\documentclass[a4paper,11pt]{article}
\pdfoutput=1 

\usepackage{jheppub} 

\usepackage[T1]{fontenc} 

\usepackage{dcolumn}
\usepackage{bm}
\usepackage{graphicx}
\usepackage{slashed}
\usepackage{appendix}
\usepackage{color}
\usepackage{tabu}
\usepackage{amsmath}
\usepackage{amsfonts}
\usepackage{makecell}
\usepackage{array,mathtools,amssymb,booktabs}

\title{\boldmath Strong cosmic censorship for the Dirac field in the higher dimensional Reissner-Nordstrom--de Sitter black hole}

\author[a]{Xiaoyi Liu}
\author[b]{Stijn Van Vooren}
\author[a,b]{Hongbao Zhang}
\author[a]{Zhen Zhong}

\affiliation[a]{Department of Physics, Beijing Normal University, Beijing 100875, China}
\affiliation[b]{ Theoretische Natuurkunde, Vrije Universiteit Brussel,
	and The International Solvay Institutes,\\Pleinlaan 2, B-1050 Brussels, Belgium}

\emailAdd{xiaoyiliu@mail.bnu.edu.cn}
\emailAdd{Stijn.Van.Vooren@vub.be}
\emailAdd{hzhang@vub.ac.be}
\emailAdd{zhenzhong@mail.bnu.edu.cn}

\abstract{We investigate the strong cosmic censorship for the Dirac field in the higher dimensional Reissner-Norstrom--de Sitter black hole. To achieve this, we first use the conformal transformation trick to massage the Dirac equation to a pair of coupled equations in a meticulously chosen orthonormal basis and derive the criterion on the quasinormal modes for the violation of the strong cosmic censorship, which turns out to be independent of the spacetime dimension. Then we apply the Crank-Nicolson method to evolve our Dirac equation in the double null coordinates and extract the low-lying quasinormal modes from the evolution data by the Prony method. It is shown for the spacetime dimension $D=4, 5, 6$ under consideration that although the strong cosmic censorship is violated by the perturbation from the neutral Dirac field in the near-extremal black hole, the strong cosmic censorship can be restored when the charge of the Dirac field is increased beyond a critical value. The closer to the extremal limit the black hole is, the larger the critical charge of the Dirac field is.}

\begin{document} 
\maketitle
\flushbottom
\section{Introduction}
In support of the predictive power of general relativity in the presence of the inevitable singularity formed by gravitational collapse, Penrose proposed his cosmic censorship conjectures a long time ago. One is called the weak cosmic censorship, which states that the formed singularity is generically hidden inside of the black hole event horizon such that the determinism is well preserved outside of the black hole. The other is called the strong cosmic censorship (SCC), which asserts that the formed singularity is generically either spacelike or null such that the determinism can hold up to the singularity, where the spacetime ends. With this in mind, the timelike singularity present in the eternal charged or rotating black hole is believed to be an illusion. To put it another way, a generic perturbation will make the Cauchy horizon (CH) singular such that one cannot extend beyond the would be CH, where the spacetime is terminated. Such a belief comes mainly from the observation that the perturbation will be blueshifted while traveling along the CH, which cannot be overshadowed by the power law decay tails outside of the black hole.

But the situation varies in the presence of a positive cosmological constant, because for the black hole in de Sitter space, it is the quasinormal modes that control the late time behavior of the perturbation outside of the black hole. Thus the validity of the SCC, namely the inextendibility of the CH depends delicately on the competition between the blueshift amplification along the CH and the exponential decay behavior outside of the black hole. In particular, for a fixed near-extremal charged Reissner-Nordstrom de Sitter (RNdS) black hole, neither the neutral scalar nor the gravito-electromagnetic perturbations can lead to an inextendible CH\cite{CCDHJ1,DRS1,LZCCN}. However, the formation of the RNdS black hole entails the participation of the charged matter fields\cite{Hod2}. As shown in \cite{CCDHJ2,MTWZZ,DRS2,Hod3}, the perturbation from the charged scalar field can preserve the SCC by making the CH inextendible except for a highly near-extremal RNdS black hole, where the SCC holds only when the scalar field is appropriately charged. On the other hand, for the perturbation from the charged Dirac field, the SCC can be respected when the Dirac field is sufficiently charged, albeit the existence of the subtle wiggles for a highly near-extremal RNdS black hole\cite{GJWZZ,Destounis}.

Generically, the physics in a higher dimensional spacetime has a richer dynamics, for example, it is shown that the RNdS black hole is gravitational unstable for the large value of the cosmological constant and electric charge in the spacetime dimension $D\ge7$\cite{KZ1,CLM,KZ2}.   So among other follow-up works\cite{DERS,Hod1,Gwak1,GG,Gwak2,Etesi,Rahman,GLKW,GGWW,DFMP}, the test of the SCC has been generalized to the scalar field in the higher dimensional RNdS black hole\cite{RCSS,LTDWPZ}. Along this line, the purpose of this paper is to investigate the validity of the SCC for the charged Dirac field in the higher dimensional RNdS black hole. To this end, we are forced to give up on the Newman-Penrose formalism, which is amenable solely to the calculation for the four dimensional Dirac field. Instead we simplify the Dirac equation in the higher dimensional RNdS background into a pair of coupled equations in Section \ref{II} by resorting to the conformal transformation trick and meticulous choice of the orthonormal basis. In Section \ref{III}, we then derive the criterion for the validity of the SCC in terms of the imaginary part of the quasinormal modes of the Dirac field. In Section \ref{V}, we present our numerical result on the validity of the SCC, where we apply the Crank-Nicolson method to evolve the Dirac equation in the double null coordinates and use the Prony method to extract the low-lying quasinormal modes from the evolution data. We conclude our paper in the last section.

The signature we take here is $(-,+,+,\cdots,+)$.

\section{Dirac Equation in the $D$-dimensional RNdS black hole\label{II}}

Let us start from the real action of the $D$-dimensional Dirac field with the mass $m$ and the charge $q$ in a charged curved spacetime
\begin{equation}
S=i\int \mathrm{d}^{D}x \sqrt{-g}\ [ \bar{\psi}(\overrightarrow{\slashed D} -m)\psi -\bar{\psi}(\overleftarrow{\slashed D}+m) \psi].
\end{equation}
Here $\bar{\psi}=\psi^\dagger \gamma^{0}$, $\overrightarrow{\slashed D}=\gamma^a\overrightarrow{D}_a=\gamma^a(\overrightarrow{\nabla}_a-iqA_a)= \gamma^{a}(\overrightarrow{\partial}_{a}+\frac{1}{4}\omega_{IJa}\gamma^{IJ}-iqA_{a})$ and
$\overleftarrow{\slashed D}=\gamma^a\overleftarrow{D}_a=\gamma^a(\overleftarrow{\nabla}_a+iqA_a)= \gamma^{a}(\overleftarrow{\partial}_{a}-\frac{1}{4}\omega_{IJa}\gamma^{IJ}+iqA_{a})$ with  $\gamma^a=\gamma^I e_{I}^a$, the spin connection $\omega_{IJa}=e_{Ib}\nabla_{a}e_{J}^b$, and $\gamma^{IJ}=\frac{1}{2}[\gamma^I,\gamma^J]$, where $\{\gamma^I\}$ are Gamma matrices satisfying $\{\gamma^I,\gamma^J\}=2\eta^{IJ}$ and $\{e^a_I\}$ form the orthonormal basis.

The energy momentum tensor for our Dirac field can be obtained by the variation of the action with respect to the metric as
\begin{equation}\label{em}
T_{ab}=\frac{2}{\sqrt{-g}}\frac{\delta S}{\delta g^{ab}}=i [ \bar{\psi}\gamma_{(b}\overrightarrow{D}_{a)}\psi -\bar{\psi}\gamma_{(b}\overleftarrow{D}_{a)}\psi ]
.
\end{equation}
On the other hand, by the variation of the above action with respect to $\psi$, we obtain the Dirac equation as
\begin{equation}
(\overrightarrow{\slashed D}-m)\psi=0,
\end{equation}
which is invariant under the gauge transformation $(A\rightarrow A+d\lambda,\psi\rightarrow e^{iq\lambda}\psi)$.

In what follows, we shall  derive an explicit expression of the above Dirac equation in a $D$-dimensional RN black hole  with the mass and charge parameter $M$ and $Q$ in the de Sitter space of radius $L$ 
\begin{equation}\label{originalmetric}
d s^2=-f(r)d t^2+\frac{d r^2}{f(r)}+r^2d\Sigma_{D-2}^2,\quad A_a=-\sqrt{\frac{D-2}{2(D-3)}}\frac{Q}{r^{D-3}}(dt)_a
\end{equation}
where the blackening factor is given by 
\begin{equation}\label{blf}
f(r)=1-\frac{M}{r^{D-3}}+\frac{Q^2}{r^{2(D-3)}}-\frac{r^2}{L^2},
\end{equation}
and 
\begin{equation}
d\Sigma_{D-2}^2=(d\phi^1)^2+\sum_{i=2}^{D-2}\prod_{j=1}^{i-1}\sin^2\phi^j(d\phi^i)^2
\end{equation}
is the line element of a $(D-2)$-dimensional unit sphere with the volume $\omega_{D-2}=\frac{2\pi^{\frac{D-1}{2}}}{\Gamma (\frac{D-1}{2})}$. The cosmological constant, the ADM mass and the charge of the black hole are given by
\begin{equation}
\Lambda=\frac{(D-1)(D-2)}{2L^2},\quad \mathcal{M}=\frac{D-2}{16\pi}\omega_{D-2} M, \quad \mathcal{Q}=\frac{\sqrt{2(D-2)(D-3)}}{8\pi}\omega_{D-2}
Q.\end{equation}
The moduli space for such black holes can be parametrized by $(\frac{\Lambda}{\Lambda_{max}},\frac{Q}{Q_{max}})$, where $\Lambda_{max}$ is the maximal cosmological constant so that $f(r)$ can have three positive real roots, giving rise to the cosmological horizon $r_c$, the black hole horizon $r_+$, and the CH $r_-$ individually, while $Q_{max}$  corresponds to the charge of the extremal black hole with $r_+=r_-$ for a given $\Lambda$.

Next by generalizing the conformal transformation \cite{GR,GRS,RS} to include the electric charge and electric potential as follows

\begin{equation}\label{cf1}
\tilde{g}_{ab}=\Omega^2 g_{ab}, \tilde{\gamma}^I=\gamma^I, \tilde{e}_I^a=\Omega^{-1}e_I^a, \tilde{q}=q, \tilde{A}_a=A_a, \tilde{m}=\Omega^{-1}m, \tilde{\psi}=\Omega^{-\frac{D-1}{2}}\psi,
\end{equation}
we have
\begin{equation}\label{cf2}
(\overrightarrow{\tilde{\slashed D}}-\tilde{m})\tilde{\psi}=\Omega^{-\frac{D+1}{2}}(\overrightarrow{\slashed D}-m)\psi.
\end{equation}
With this in mind, we can consider the  equivalent Dirac equation in the following conformally transformed background
\begin{equation}\label{originalmetric}
d\tilde{s}^2=\frac{1}{r^2} d s^2=-\frac{f}{r^2}d t^2+\frac{d r^2}{fr^2}+d\Sigma_{D-2}^2.
\end{equation}
To proceed, we choose the following orthonormal basis
\begin{equation}
\begin{split}
&\tilde{e}_{t}^a =\frac{r}{2\sqrt{2}}\left[\frac{2+f}{f} \left( \frac{\partial}{\partial t}\right)^a+(2-f)\left(\frac{\partial}{\partial r}\right)^a\right],\\
&\tilde{e}_{r}^a =\frac{r}{2\sqrt{2}}\left[\frac{2-f}{f} \left( \frac{\partial}{\partial t}\right)^a+(2+f)\left(\frac{\partial}{\partial r}\right)^a\right],\\
&\tilde{e}_1^a= \left( \frac{\partial}{\partial \phi^{1}}\right)^a, \quad
\tilde{e}_{i}^a =\frac{1}{\prod_{j=1}^{i-1}\sin\phi^j} \left( \frac{\partial}{\partial \phi^{i}}\right)^a
\end{split}
\end{equation}
with  $i=2,\cdots, D-2$. In addition, for even $D$, we choose the Gamma matrices in terms of Pauli matrices as\cite{L}
\begin{equation}
{\gamma}^{t} =i\sigma_{1} \otimes \mathbb{I}_{2^{(D-2) / 2}},\quad
{\gamma}^{r} = \sigma_{2} \otimes \mathbb{I}_{2^{(D-2) / 2}},\quad
{\gamma}^{i} =\sigma_{3} \otimes \hat{\gamma}^{i}
\end{equation}
with $ i=1,\cdots,D-2$, where
\begin{equation}
\hat{\gamma}^{2n-1}=\sigma_3^{\otimes(n-1)}\otimes\sigma_1\otimes \mathbb{I}_{2^{(D-2) / 2-n}}, \quad \hat{\gamma}^{2n}=\sigma_3^{\otimes(n-1)}\otimes\sigma_2\otimes \mathbb{I}_{2^{(D-2) / 2-n}}
\end{equation}
with $n=1,\cdots,(D-2)/2$. Similarly, for odd $D$, the Gamma matrices are chosen as follows
\begin{equation}
{\gamma}^{t} =i\sigma_{1} \otimes \mathbb{I}_{2^{(D-3) / 2}},\quad
{\gamma}^{r} = \sigma_{2} \otimes \mathbb{I}_{2^{(D-3) / 2}},\quad
{\gamma}^{i} =\sigma_{3} \otimes \hat{\gamma}^{i}
\end{equation}
with $ i=1,\cdots,D-2$, where
\begin{equation}
\hat{\gamma}^{2n-1}=\sigma_3^{\otimes(n-1)}\otimes\sigma_1\otimes \mathbb{I}_{2^{(D-3) / 2-n}}, \quad \hat{\gamma}^{2n}=\sigma_3^{\otimes(n-1)}\otimes\sigma_2\otimes \mathbb{I}_{2^{(D-3) / 2-n}},\quad \hat{\gamma}^{D-2}=\sigma_3^{\otimes(D-3)/2}
\end{equation}
with $n=1,\cdots,(D-3)/2$. It is noteworthy that $ \{\hat{\gamma}^i\}$ denote the Gamma matrices for  a $(D-2)$-dimensional space with the signature $(+,\cdots,+)$.

Then it is not hard to show that the aforementioned choice gives rise to
\begin{equation}\label{D}
\tilde{\slashed \nabla} =\left(i\sigma_1\tilde{e}_t^a \nabla_{a}^{(2)}+\sigma_2\tilde{e}_r^a \nabla_{a}^{(2)}\right) \otimes I+{\sigma_{3} \otimes\left(\hat{\gamma}^{i}\tilde{e}_i^a \nabla_{a}^{(D-2)}\right)} =\slashed \nabla_{2} \otimes {I}+ \sigma_{3} \otimes \slashed \nabla_{\Sigma},
\end{equation}
where  $\slashed \nabla_{2}$ is the Dirac operator associated with the two-dimensional spacetime
\begin{equation}\label{2dst}
d s_{2}^{2}=-\frac{f}{r^{2}} d t^{2}+\frac{d r^{2}}{fr^2},
\end{equation}
and $\slashed \nabla_{\Sigma}$  is the Dirac operator on $(D-2)$-dimensional unit sphere $\Sigma_{D-2}$. Accordingly, we are motivated to make the following separation of variables for our Dirac field as
\begin{equation}\label{decom}
\tilde{\psi}\left(r, t, \phi^{i}\right)=\varphi(r, t) \otimes \chi\left(\phi^{i}\right)
\end{equation}
where $\varphi(r,t)$ is a two-component spinor field on the spacetime (\ref{2dst}) and $\chi(\phi^{i})$ is the eigen spinor fields of the Dirac operator in the unite sphere $\Sigma_{D-2}$, that is to say,
\begin{equation}
\overrightarrow{\slashed \nabla_{\Sigma}} \chi=K \chi
\end{equation}
with $K=\pm i(l+\frac{D-2}{2}),\ l=0,1,2,\cdots$\cite{CH}. Whence the Dirac equation reduces to
\begin{equation}
\overrightarrow{\slashed D_{2}}\varphi=(-K\sigma_{3}+rm)\varphi,
\end{equation}
where we have used the $\tilde{m}=rm$ associated with the previous conformal transformation (\ref{originalmetric}).
Let $\varphi=(\sqrt{\frac{r}{2}}\tilde{\varphi}_1,\sqrt{\frac{r}{f}}\tilde{\varphi}_2)$,  then we end up with an explicit expression for our Dirac equation as
\begin{equation}\label{partialori}
\begin{split}
(\partial_{t}-iqA_t) \tilde{\varphi}_{2}-f (\partial_{r}-iqA_r) \tilde{\varphi}_{2}&=i\sqrt{f}\left(\frac{K}{r}-m \right) \tilde{\varphi}_{1}, \\
(\partial_{t}-iqA_t) \tilde{\varphi}_{1}+f (\partial_{r}-iqA_r) \tilde{\varphi}_{1}&=-i\sqrt{f}
\left(\frac{K}{r}+m \right) \tilde{\varphi}_{2},
\end{split}
\end{equation}
where we have used the non-vanishing spin connection $\omega_{tra}=\frac{2f-f'r}{2r}(dt)_a+\frac{f'}{2f}(dr)_a$.

\section{Quasinormal modes and strong cosmic censorship\label{III}}
With the seperation $\tilde{\varphi}=\tilde{R}(r)e^{-i\omega t}$,  the above equation can be written as
\begin{equation}\label{finaldirac}
\begin{split}
-i[\omega-\Phi(r)]\tilde{R}_{2}-f(r)(\partial_r-iqA_r) \tilde{R}_{2}&=i\sqrt{f}\left(\frac{K}{r}-m \right) \tilde{R}_{1},\\
-i[\omega-\Phi(r)]\tilde{R}_1+f(r)(\partial_r-iqA_r)\tilde{R}_1&=-i\sqrt{f}
\left(\frac{K}{r}+m \right) \tilde{R}_{2},
\end{split}
\end{equation}
which gives rise to  two independent asymptotic solutions
\begin{equation}
\begin{split}
&\tilde{R}_1\sim\sqrt{f}e^{-i[\omega-\Phi(r_h)]r_*},\quad\tilde{R}_2\sim e^{-i[\omega-\Phi(r_h)]r_*},\\
&\tilde{R}_1\sim e^{i[\omega-\Phi(r_h)]r_*},\quad\tilde{R}_2\sim\sqrt{f}e^{i[\omega-\Phi(r_h)]r_*}
\end{split}
\end{equation}
near any horizon $r_h$ with the surface gravity defined as $\kappa_h=\frac{1}{2}|f'(r_h)|$, where $\Phi(r)=\sqrt{\frac{D-2}{2(D-3)}}\frac{qQ}{r^{D-3}}$ is the electric potential energy and $r_*=\int\frac{dr}{f}$ is the tortoise coordinate. By imposing the outgoing boundary condition near the cosmological horizon and the ingoing boundary condition near the black hole event horizon, one can obtain a discrete spectrum of quasinormal modes, which have the symmetry $\omega\rightarrow -\bar{\omega}$ under $q\rightarrow -q$.

However, these quasinormal modes propagating into the black hole will generically have both the outgoing mode
\begin{equation}
\varphi_1\sim e^{-i\omega u},\quad \varphi_2\sim e^{-i\omega u}
\end{equation}
and the ingoing mode
\begin{equation}
\varphi_1\sim(r-r_-)^{\frac{1}{2}+\frac{i[\omega-\Phi(r_-)]}{\kappa_-}}e^{-i\omega u},\quad \varphi_2\sim (r-r_-)^{-\frac{1}{2}+\frac{i[\omega-\Phi(r_-)]}{\kappa_-}}e^{-i\omega u}
\end{equation}
near the CH, where we have made a gauge transformation $d\lambda=\sqrt{\frac{D-2}{2(D-3)}}\frac{Q}{r^{D-3}}dr_*$ and moved onto the outgoing coordinates $(u,r)$ with the retarded Eddington time $u=t-r_*$ such that the background can be analytically continued across the CH. Note that 
\begin{equation}
\tilde{e}_t^a=\frac{r}{\sqrt{2}}\left[\left(\frac{\partial}{\partial u}\right)^a+\frac{2-f}{2}\left(\frac{\partial}{\partial r}\right)^a\right], \quad \tilde{e}_r^a=\frac{r}{\sqrt{2}}\left[-\left(\frac{\partial}{\partial u}\right)^a+\frac{2+f}{2}\left(\frac{\partial}{\partial r}\right)^a\right]
\end{equation}
as well as the non-vanishing spin connection $\omega_{tra}=\frac{2f-f'r}{2r}(du)_a+\frac{1}{r}(dr)_a$ are also smoothly across the CH, so the potential non-smoothness of our energy momentum tensor (\ref{em}) comes only from such terms as $\bar{\varphi}_i\varphi_j$ and $\bar{\varphi}_i\partial_\mu\varphi_j$ with $\varphi_i$ either the ingoing or outgoing mode, where the dominant term is given obviously by 
$\bar{\varphi}_2\partial_r\varphi_2$ with $\varphi_2$
 the ingoing mode. Speaking specifically, near the CH the energy momentum tensor from this term behaves as 
\begin{equation}
T_{ab}\sim (r-r_-)^{\frac{2i[\omega-\Phi(r_-)]}{\kappa_-}-2},
\end{equation}
which can induce a weak solution to Einstein equation if and only if it is integrable near the CH, i.e., 
\begin{equation}
\beta\equiv -\frac{\mathrm{Im}(\omega)}{\kappa_-}> \frac{1}{2}
\end{equation}
for all quasinormal modes. In other words, if there exists a quasinormal mode with $\beta\le\frac{1}{2}$, then one can not extend beyond the CH such that the SCC is preserved. Obviously, to examine the validity of the SCC, we only need to focus on the lowest-lying quasinormal mode.  Note that the charge to mass
ratio of the electron in our universe is very large, so below we shall specialize in the massless case.

\section{Numerical results\label{V}}

\subsection{Numerical Scheme}
In this section, we would like to extract the low-lying quasinormal modes by the time domain analysis of the numerical solution to the Dirac equation in the double null coordinates $(u,v)$
\begin{equation}\label{dirac}
(\partial_{u}-iqA_u) \tilde{\varphi}_{2}=i\frac{\sqrt{f}}{2}\left(\frac{K}{r}-m \right) \tilde{\varphi}_{1}, \quad
(\partial_{v}-iqA_v) \tilde{\varphi}_{1}=-i\frac{\sqrt{f}}{2}
\left(\frac{K}{r}+m \right) \tilde{\varphi}_{2},
\end{equation}
which has been shown to be naturally suitable for the numerical evolution by the Crank-Nicolson method along the double null directions with the advanced Eddington time $v=t+r_*$\cite{GJWZZ}. $r$ in the above equation can be solved out through
\begin{equation}
r_*=\sum_{i=1}^{2(D-2)} \frac{\ r_i^{2(D-3)}\ln (r-r_i)}{g'(r_i)}+C,
\end{equation}
where $g(r)=r^{2(D-3)}f(r)$, $r_i$s are the zeros of $g(r)$, and the undetermined coefficient $C$ can be specified such that $r_*=0$ when $r=\frac{r_+ + r_c}{2}$.

To set off such a numerical evolution, we impose the initial conditions on the initial double null surfaces as
\begin{equation}
\tilde{\varphi_2}(0, v) =\frac{1}{\sqrt{2 \pi} w_{1}} e^{-\frac{\left(v-v_{c}\right)^{2}}{2 w_{1}^{2}}}, \quad \tilde{\varphi_2}(u, 0) =\frac{1}{\sqrt{2 \pi} w_{2}} e^{-\frac{\left(u-u_{c}\right)^{2}}{2 w_{2}^{2}}}, 
\end{equation}
and $\tilde{\varphi_1}$ can be obtained by solving the Dirac equation (\ref{dirac}) on these surfaces. In addition, we also make a gauge transformation such that the electric potential reads
\begin{equation}
A_{a}=-\sqrt{\frac{D-2}{2(D-3)}}\ \frac{Q\left(r-r_{+}\right)}{r\left(r_{c}-r_{+}\right)}(d u)_{a}+\sqrt{\frac{D-2}{2(D-3)}}\ \frac{Q\left(r-r_{c}\right)}{r\left(r_{c}-r_{+}\right)}(d v)_{a}.
\end{equation}
Then we can check the validity of the SCC by examining the low-lying quasinormal modes extracted from the evolution data through the Prony method\cite{BCGS}. In what follows, we shall work with the units in which the ADM mass $\mathcal{M}=1$. 
In addition, due to our limited computational resources, below we shall narrow down our discussions to the massless Dirac field in the $D=4, 5, 6$-dimensional near-extremal RNdS black hole with $\frac{\Lambda}{\Lambda_{max}}=0.27$ and $\frac{Q}{Q_{max}}=0.996, 0.999$.

\subsection{Relevant Results}

\begin{table}[htbp]
\centering
\begin{tabular}{cccc}
\toprule
& $D=4$ & $D=5$ & $D=6$\\ \midrule$\ l=0\ $ & \makecell{
 $(\pm2.1322,-0.7546)$} & \makecell{
 $(\pm2.6335,-1.0223)$} & \makecell{$(0,-1.0931)$\\$(\pm2.9677,-1.1260)$} \\ \midrule$\ l=1\ $ & $(\pm4.3806,-0.7621)$ & $(\pm4.5306,-1.0345)$ & $(\pm4.5814,-1.1388)$ \\ \midrule$\ l=2\ $ & $(\pm6.6031,-0.7633)$ & $(\pm6.3977,-1.0377)$ & $(\pm6.1696,-1.1430)$ \\ \midrule
 WKB & $(\ \ \ \quad\ \ \ \ \ ,-0.7643 )$ & $(\ \ \ \quad\ \ \ \ \ ,-1.0410 )$ & $(\ \ \ \ \ \quad\ \ \ , -1.1486)$ \\\bottomrule
\end{tabular}
\caption{The low-lying spectrum of quasinormal modes $\frac{\omega}{\kappa_-}$ with $q=0$, $\frac{\Lambda}{\Lambda_{max}}=0.27$, and $\frac{Q}{Q_{max}}=0.996$.}\label{QNM}\end{table}

\begin{table}[htbp]
\centering
\begin{tabular}{cccc}\toprule
& $D=4$ & $D=5$ & $D=6$\\ \midrule$\ l=0\ $ & \makecell{$(0,-1.4364)$ \\ $(\pm4.7394,-1.6647)$} & \makecell{$(0,-1.2254)$ \\ $(0,-2.1699)$ \\ $(\pm5.6824,-2.2016)$} & \makecell{$(0,-1.1545)$ \\ $(0,-2.1042)$ \\ $(\pm6.3430,-2.4044)$} \\ \midrule$\ l=1\ $ & $(\pm9.7482,-1.6791)$ & \makecell{$(0,-1.7284)$ \\ $(\pm9.7805, -2.2263)$} & \makecell{$(0,-1.4948)$ \\ $(\pm9.7944, -2.4308)$} \\ \midrule $\ l=2\ $ & $(\pm14.6969,-1.6814)$ & \makecell{$(0,-2.2321)$ \\ $(\pm13.8131, -2.2328)$} & \makecell{$(0,-1.8355)$ \\ $(\pm13.1909, -2.4395)$} \\ \midrule WKB & $(\ \ \ \quad\ \ \ \ \ ,-1.6832)$ & $(\ \ \ \quad\ \ \ \ \ , -2.2396)$ & $(\ \ \ \ \ \quad\ \ \ , -2.4510)$ \\\bottomrule
\end{tabular}
\caption{The low-lying spectrum of quasinormal modes $\frac{\omega}{\kappa_-}$ with $q=0$, $\frac{\Lambda}{\Lambda_{max}}=0.27$, and $\frac{Q}{Q_{max}}=0.999$.}
\label{MNQ}
\end{table}

\begin{figure}
	\centering
	\includegraphics[width=10.0cm]{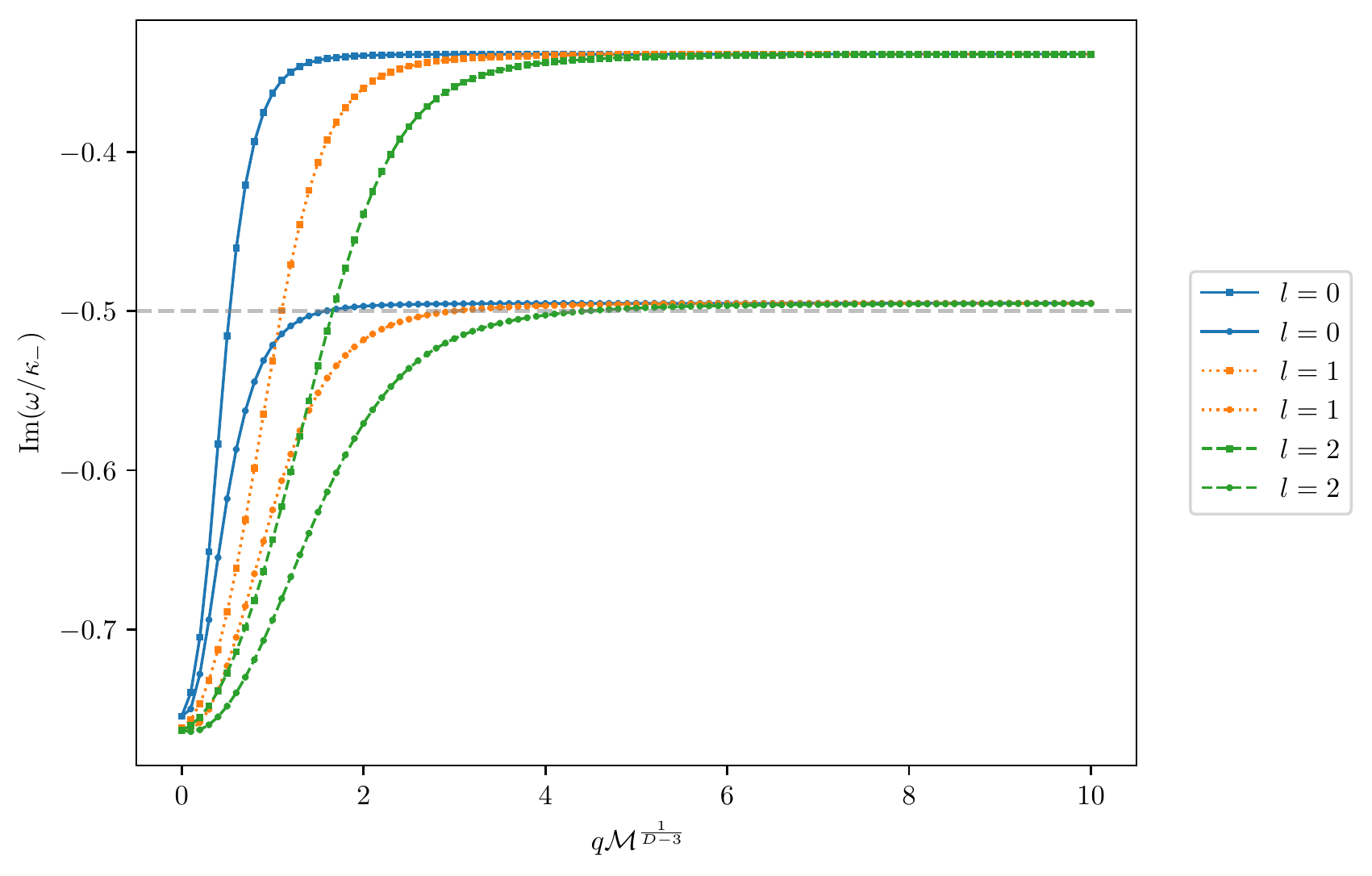}
	\includegraphics[width=10.0cm]{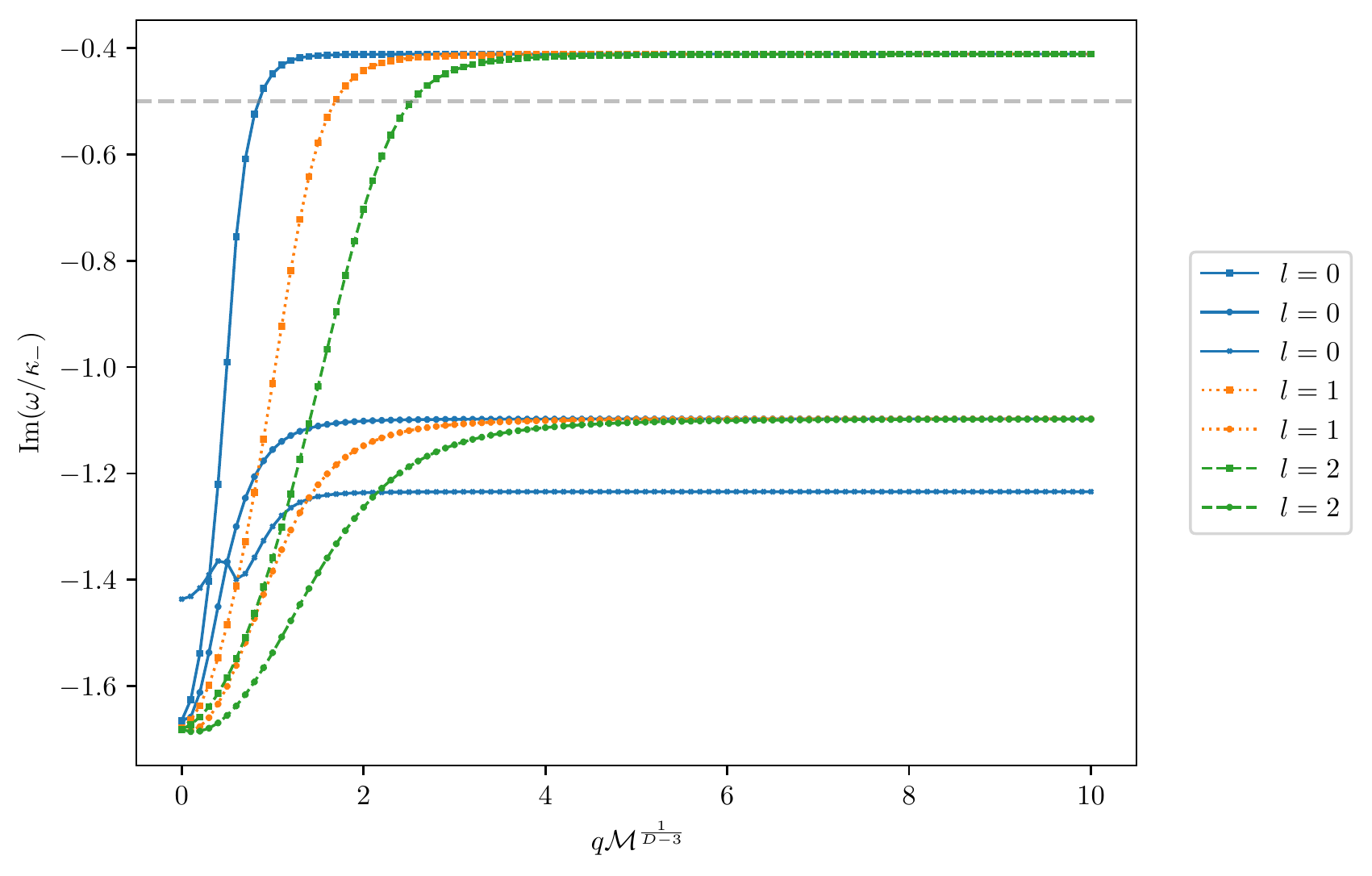}
		\caption{The imaginary part of the low-lying quasinormal modes for $\frac{\Lambda}{\Lambda_{max}}=0.27$  in $D=4$, where the top is for and $\frac{Q}{Q_{max}}=0.996$ and the bottom is for $\frac{Q}{Q_{max}}=0.999$.}\label{d4}
\end{figure}

\begin{figure}
	\centering
	\includegraphics[width=10.0cm]{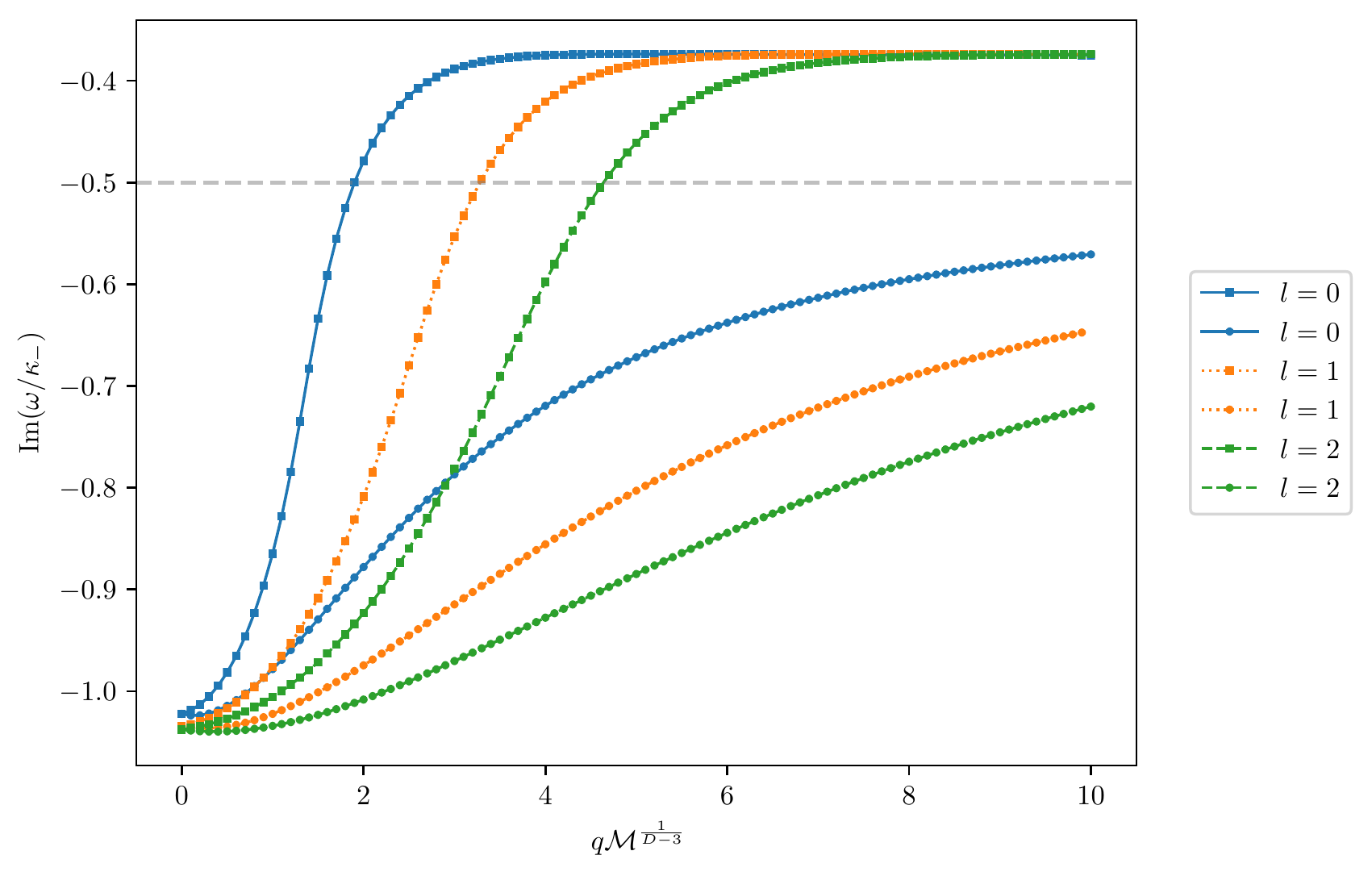}
	\includegraphics[width=10.0cm]{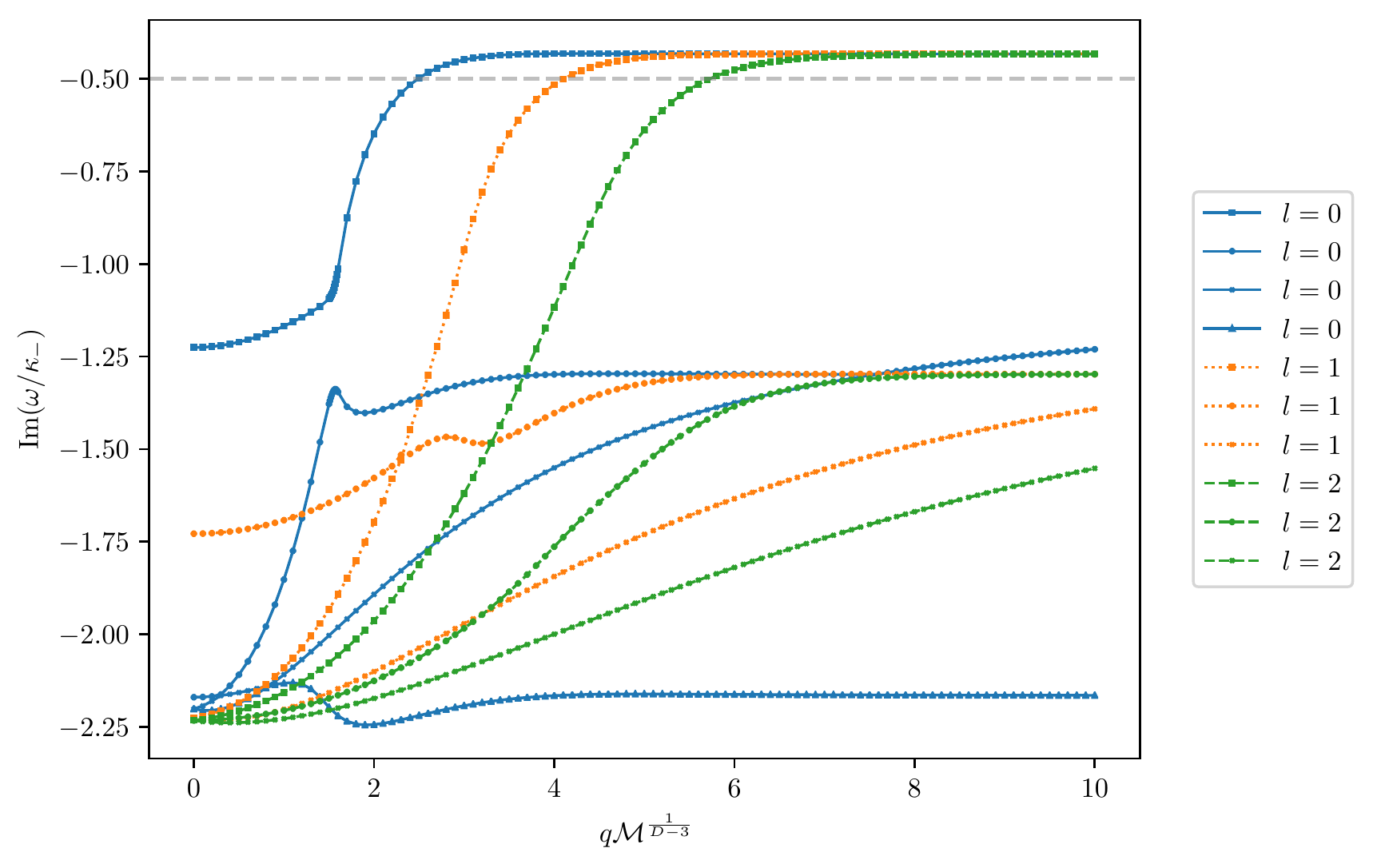}	
	\caption{The imaginary part of the low-lying quasinormal modes for $\frac{\Lambda}{\Lambda_{max}}=0.27$ in $D=5$, where the top is for and $\frac{Q}{Q_{max}}=0.996$ and the bottom is for $\frac{Q}{Q_{max}}=0.999$.}\label{d5}
\end{figure}

\begin{figure}
	\centering
	\includegraphics[width=10.0cm]{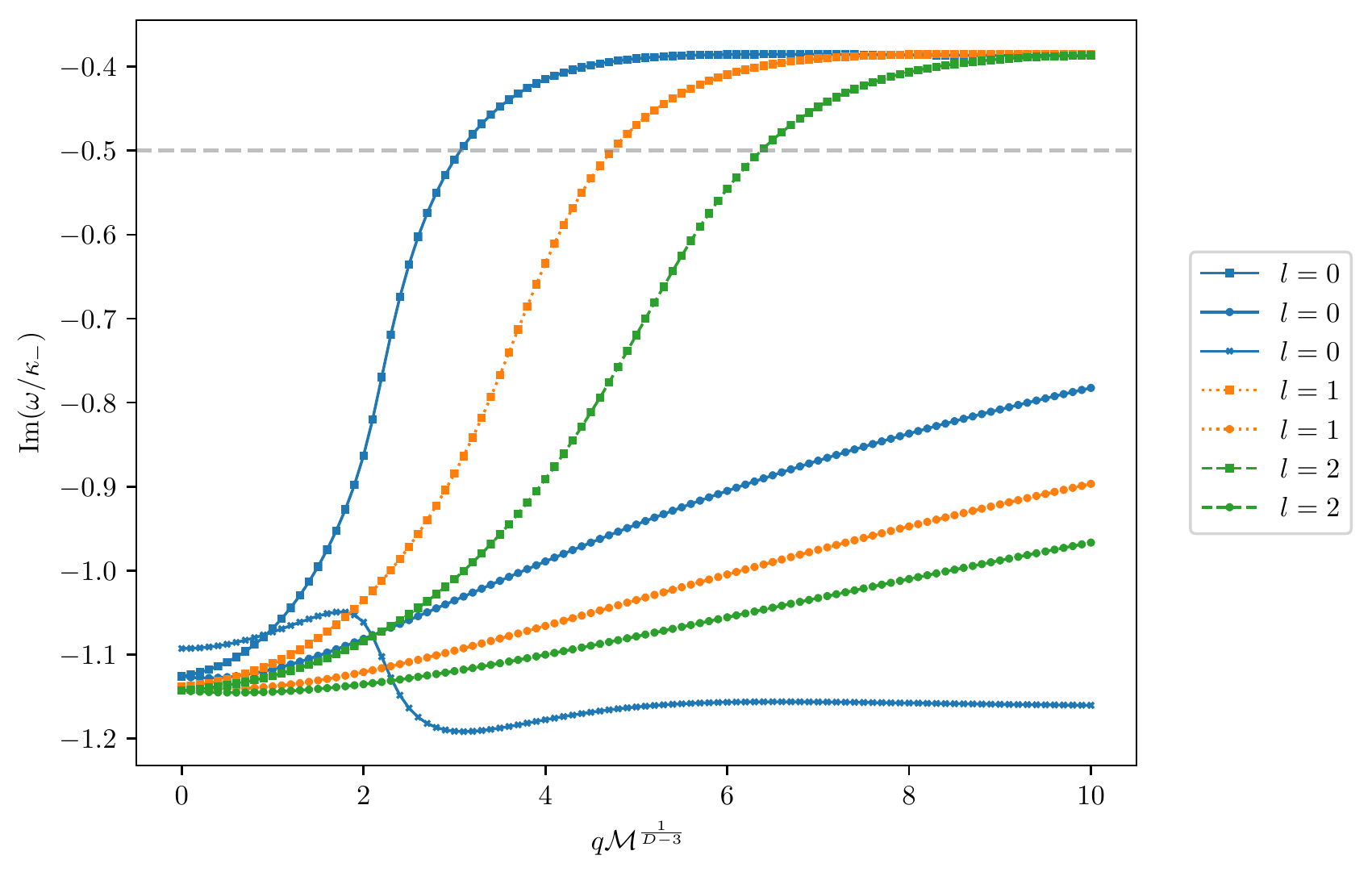}
	\includegraphics[width=10.0cm]{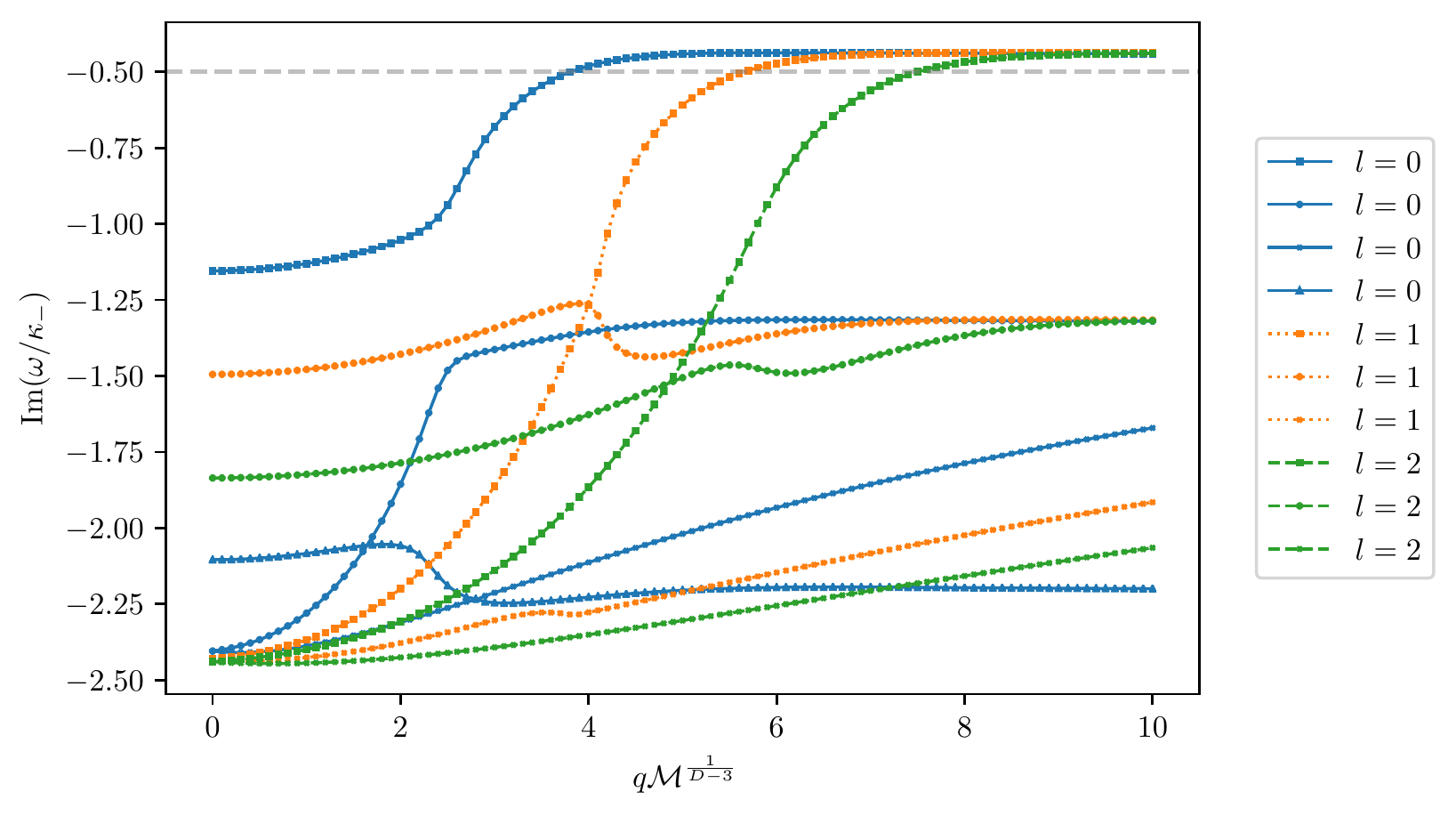}
		\caption{The imaginary part of the low-lying quasinormal modes for $\frac{\Lambda}{\Lambda_{max}}=0.27$ in $D=6$, where the top is for and $\frac{Q}{Q_{max}}=0.996$ and the bottom is for $\frac{Q}{Q_{max}}=0.999$.}\label{d6}
\end{figure}
First we list the low-lying quasinormal modes for the neutral Dirac field in different dimensions in Table \ref{QNM} and Table \ref{MNQ}, where for the large $l$ WKB limit (See Appendix A), we only present the imaginary part of the corresponding quasinormal mode because of the divergence of the real part. As expected, the spectrum of quasinormal modes is symmetric with respect to the imaginary axis of the $\omega$-plane. In addition, we find that the dominant mode always comes from the $l=0$ mode. In particular, by comparing Table \ref{QNM} and Table \ref{MNQ} for $D=4$ and $D=5$, one can see that although for the less charged near-extremal RNdS black hole the dominant mode is from the oscillatory photon sphere mode\footnote{The photon sphere mode is so called because it can be traced back to the properties of the unstable circular null orbits at the large $l$ with its real part proportional to the angular velocity of the null orbits and its imaginary part describing how quickly the null orbits are driven away from the unstable circular orbits under the infinitesimal radial deformations.}, the purely imaginary mode will become dominant one when the black hole charge is further increased towards the extremal limit.
Furthermore, it follows from the dominant mode that the SCC is violated for the perturbations by the neutral Dirac field. 

We next demonstrate how the imaginary part of the low-lying quasinormal modes varies with the increase of the charge of the Dirac field in Figure.\ref{d4}, Figure.\ref{d5}, and Figure.\ref{d6} separately for $D=4, 5, 6$.  Among others, one can see for all the dimensions under consideration that the aforementioned symmetry of the photon sphere mode pair breaks down once the Dirac field is charged. In addition, the dominant mode for the charged Dirac field is still from the $l=0$ mode. Speaking specifically, for the less charged near-extremal black hole, the dominant mode comes solely from one of the photon sphere mode pair. For the more charged near-extremal black hole, the dominant mode is given solely by the branch arising from the aforementioned purely imaginary mode. When the near-extremal black hole is intermediately charged, the role of the dominant mode is shifted from the purely imaginary mode branch to one of the photon sphere mode pair as the charge of the Dirac field is cranked up to a certain value. Such a general pattern is consistent with the observation made for $D=4$ in \cite{Destounis}, where more details can be found. But nevertheless, for our purpose, whichever plays the dominant mode, the imaginary part of the dominant mode increases with the increase of the charge of the Dirac field.  Actually in the large $q$ limit, the dominant mode falls into the black hole family with the imaginary part approaching $-\frac{\kappa_+}{2}$ (See Appendix B). In particular, the dominant mode restores the SCC once the charge of the Dirac field is more than a certain critical value, which is increased as the black hole charge is increased.

\section{Conclusion}
To check the validity of the SCC in the higher dimensional RNdS black hole under the perturbation of the Dirac field, we first apply the conformal transformation trick such that the Dirac equation can be reduced to a pair of coupled equations in the well chosen orthonormal basis, which is smoothly across the CH. With this, we further derive the criterion on the quasinormal modes for the violation of the SCC, which is exactly the same for all dimensions. Then we mainly employ the Crank-Nicolson method to evolve the involved Dirac equation in the double null coordinates and extract the low-lying quasinormal modes from the resulting evolution data by the Prony method for $D=4, 5, 6$. As a result, although the SCC is violated for the perturbation of the neutral Dirac field in the near-extremal RNdS black hole, the SCC can be recovered once one charges the Dirac field beyond a critical value, which becomes larger when the black hole is closer to the extremal limit. We believe that such a pattern is also true for any other higher dimensional RNdS black hole.

It is noteworthy that for the highly extremal higher dimensional RNdS black hole, the non-perturbative wiggles may show up in the large $q$ regime, causing the violation of the SCC over there as in $D=4$\cite{Destounis}. Although we are unable to probe such a phenomenon by numerics due to our limited computational resources, we believe that the SCC will be recovered again in the sufficiently large $q$ regime due to the fact $\kappa_+<\kappa_-$ proven in Appendix C.

\acknowledgments

 This work is partially supported by NSFC with Grant 
No.11675015  and No.11775022  as well as by FWO-Vlaanderen through the
project G020714N, G044016N, and G006918N.  XL is supported by the National Research Fund for Talented Undergraduates. HZ is an individual FWO Fellow supported by 12G3515N and by the
Vrije Universiteit Brussel through the Strategic Research Program "High-Energy Physics".

\appendix
\section{ WKB approximation for the large $l$ limit}
In order to obtain the WKB approximation of the quasinormal modes for the large $l$ limit, we like to define $Z_\pm=\tilde{R}_2\pm\tilde{R}_1$, $\frac{d\hat{r}_*}{dr_*}=1-\frac{\Phi}{\omega}$, and $W=(1-\frac{\Phi}{\omega})^{-1}\frac{\sqrt{f}}{r}\left(l+\frac{D-2}{2}\right)$. As a result,  the massless Dirac equation (\ref{finaldirac}) can be rewritten as
\begin{equation}\label{susy}
\left[\frac{d}{d\hat{r}_*}\pm W(r)\right]Z_\pm=-i\omega Z_\mp,
\end{equation}
which can be further expressed as a pair of decoupled equations as
\begin{equation}
\left(\frac{d^2}{d\hat{r}_*^2}+\omega^2\right)Z_\pm=V_\pm(r)Z_\pm
\end{equation}
with $V_\pm=W^2\mp\frac{dW}{d\hat{r}_*}$. Note that this pair of equations lead to the exactly same spectrum of quasinormal modes due to $Z_\pm$ are related to each other through the supersymmetric relation (\ref{susy}). Therefore we can extract the low-lying quasinormal modes by applying the third order WKB approximation formula to either equation, which is generically highly accurate in the large $l$ limit\cite{IW,Iyer,Konoplya1,Konoplya2}.
\section{ WKB approximation for the large $q$ limit}
We like to obtain the quasinormal modes in the large $q$ limit by working in the ingoing coordinates $(v,r)$ with the advanced Eddington time defined as $v=t+r_*$ and the electric potential gauged as $A_a=-\sqrt{\frac{D-2}{2(D-3)}}\frac{Q}{r^{D-3}}(dv)_a$. Then it is not hard to show that with $\tilde{\varphi}=e^{-i\omega v}\hat{R}(r)$ the massless Dirac equation (\ref{partialori}) reduces to
\begin{eqnarray}
f\frac{d\hat{R}_2}{dr}+\frac{\sqrt{f}}{r}\left(l+\frac{D-2}{2}\right)\hat{R}_1&=&0,\\
f\frac{d\hat{R}_1}{dr}-2i(\omega-\Phi)\hat{R}_1+\frac{\sqrt{f}}{r}\left(l+\frac{D-2}{2}\right)\hat{R}_2&=&0,
\end{eqnarray}
which further gives the decoupled equation for $\hat{R}_2$ as
\begin{equation}\label{WKB}
4fr^2\hat{R}_2''+[4fr+2f'r^2-8ir^2(\omega-\Phi)]\hat{R}_2'-(2l+D-2)^2\hat{R}_2=0,
\end{equation}
where the prime denotes the differentiation with respect to $r$. Following \cite{DRS2}, we make the \textit{ansatz} that
\begin{equation}
\begin{aligned} \hat{R}_2 &=\left(1-\frac{r}{r_{c}}\right)^{\frac{1}{2}-i \frac{\left[\omega-\Phi\left(r_{c}\right)\right]}{\kappa_{c}}} e^{-q \psi(r)} \sum_{n=0}^{+\infty} \frac{\hat{R}_2^{(n)}(r)}{q^{n}}, \\ \omega &=\sum_{n=-1}^{+\infty} \frac{\omega^{(n)}}{q^{n}}. \end{aligned}
\end{equation}
Substituting it into Eq.(\ref{WKB}) and solving the equation order by order in $\frac{1}{q}$, to the leading order, we can get the solution
\begin{equation}
\omega_c^{(-1)}=\sqrt{\frac{D-2}{2(D-3)}}\frac{Q}{r_c^{D-3}},\quad
\omega_+^{(-1)}=\sqrt{\frac{D-2}{2(D-3)}}\frac{Q}{r_+^{D-3}},
\end{equation}
with the subscript $c$ and $+$ denoting the solution belonging to the cosmological family or the black hole family respectively. The corresponding equation for $\psi$ is given by
\begin{equation}
\begin{split}
\psi'_c&=0, \\
\psi'_+&=\sqrt{\frac{D-2}{2(D-3)}}\frac{i Q \left(2 {\kappa _c} {r_c}^{d-3} (r-{r_c}) \left(r^{d-3}-{r_+}^{d-3}\right)+r^{d-3}  \left({r_c}^{d-3}-{r_+}^{d-3}\right)f\right)}{{\kappa _c}{r_c}^{d-3} {r_+}^{d-3} r^{d-3} (r_c-r) f }.
\end{split}
\end{equation}
For our purpose, we present the solution to the next two orders as
\begin{equation}\label{family}
\begin{split}
\omega_c^{(0)}&=-\frac{i\kappa_c}{2},\quad \omega_c^{(1)}=-\frac{(D-2+2l)^2r_c^{D-4}\kappa_c}{4\sqrt{2(D-2)(D-3)}Q},\\
\omega_+^{(0)}&=-\frac{i\kappa_+}{2},\quad \omega_+^{(1)}=\frac{(D-2+2l)^2r_+^{D-4}\kappa_+}{4\sqrt{2(D-2)(D-3)}Q}.
\end{split}
\end{equation}
It is easy to see that when $D=4$, our result is identical to \cite{GJWZZ}.

\section{ Proof of $\kappa_+<\kappa_-$ for the non-extremal black hole in $D\ge 4$}

It follows from the inspection of the blackening factor (\ref{blf}) that  the derivative of $f$ must be negative at $r_-$ and positive at $r_+$. Thus we have
\begin{equation}
\kappa_+-\kappa_-=\frac{f'(r_-)+f'(r_+)}{2}.
\end{equation}
$\kappa_+<\kappa_-$ can be proven by the brute force calculation, which turns out to be a little involved. Here we would rather present a much simpler proof as follows\footnote{We are grateful to Dongyi Wei and Shiwu Yang for their sharing with us the key strategy for this simple version of proof. }.

To achieve this, let us define the functions $h(x)=1-Mx+Q^2x^2$ and $p(x)=\frac{x^{\frac{2}{3-D}}}{L^2}$ with $x=r^{3-D}$, then by the Lagrange's interpolation formula, we have
\begin{equation}
h(x)=\frac{(x-x_+)(x-x_-)}{(x_c-x_+)(x_c-x_-)}p(x_c)+\frac{(x-x_c)(x-x_-)}{(x_+-x_c)(x_+-x_-)}p(x_+)+\frac{(x-x_c)(x-x_+)}{(x_--x_c)(x_--x_+)}p(x_-)
\end{equation}
where $f(r_c)=f(r_+)=f(r_-)=0$ with $r_c>r_+>r_-$ has been used. Whence it is not hard to show
\begin{equation}
h'(x_+)=k_{c+}+k_{-+}-k_{c-}, \quad h'(x_-)=k_{c-}+k_{+-}-k_{c+}
\end{equation}
with $k_{ij}=\frac{p(x_i)-p(x_j)}{x_i-x_j}=\int_0^1dtp'[tx_i+(1-t)x_j]$.  With this, we further have
\begin{eqnarray}
\kappa_+-\kappa_-&&=\frac{3-D}{2}[r_+^{2-D}(k_{c+}+k_{-+}-k_{c-}-p'(x_+))+r_-^{2-D}(k_{c-}+k_{+-}-k_{c+}-p'(x_-))]\nonumber\\
&&=\frac{3-D}{2}[r_+^{2-D}(F(x_c)-F(x_+))+r_-^{2-D}(F(x_-)-F(x_c))]\nonumber\\
&&=\frac{3-D}{2}[r_-^{2-D}(F(x_-)-F(x_+))+(r_+^{2-D}-r_-^{2-D})(F(x_c)-F(x_+))],
\end{eqnarray}
which is negative for $D\ge 4$ due to the fact
\begin{equation}
F(x)=\int_0^1dt \{p'[tx_++(1-t)x]-p'[tx_-+(1-t)x]\}
\end{equation}
is obviously an increasing function in $(0,+\infty)$.

\end{document}